\documentclass[12pt]{article}

\usepackage{jcappub}

\usepackage{amsmath}
\usepackage{graphicx}
\usepackage{xspace}
\usepackage{lineno}

\RequirePackage[numbers,sort&compress]{natbib}

\def\LSP{\tilde{\chi}^0_1}
\def\chip{\tilde{\chi}_1^+}

\def\ie{i.e.}
\def\eg{e.g.}
\def\cf{c.f.}
\def\sub#1{_{\lower.25ex\hbox{$\scriptstyle#1$}}}
\def\tev{\,{\rm TeV}}
\def\gev{\,{\rm GeV}}
\def\mev{\,{\rm MeV}}

\def\to{\rightarrow}
\def\ch{\ifmmode H^\pm \else $H^\pm$\fi}

\newskip\zatskip \zatskip=0pt plus0pt minus0pt
\def\matth{\mathsurround=0pt}
\def\lsim{\mathrel{\mathpalette\atversim<}}
\def\gsim{\mathrel{\mathpalette\atversim>}}
\def\atversim#1#2{\lower0.7ex\vbox{\baselineskip\zatskip\lineskip\zatskip
  \lineskiplimit
  0pt\ialign{$\matth#1\hfil##\hfil$\crcr#2\crcr\sim\crcr}}}

\def\bigd{ \ifmmode \mathcal{D} \else $\mathcal{D}$ \fi}
\def\phis{ \ifmmode \phi_s \else $\phi_s$ \fi}

\def\rphis{ \ifmmode \phi_s\times(m_{\LSP}^{2}/\langle\sigma
  v\rangle_T) \else $\phi_s\times(m_{\LSP}^{2}/\langle\sigma
  v\rangle_T)$ \fi}

\def\phiul{ \ifmmode \phi_\mathrm{UL} \else $\phi_\mathrm{UL}$ \fi}

\newcommand{\Fermi}[0]{\textit{Fermi}\xspace}

\newcommand{\bbbar}[0]{\ensuremath{b \bar b}\xspace}

\newcommand{\sigv}[0]{\ensuremath{\langle\sigma v\rangle}\xspace}
\newcommand{\sigvm}[0]{\ensuremath{\langle\sigma v\rangle_{\rm{UL}}}\xspace}
\newcommand{\tsigv}[0]{\ensuremath{\langle\sigma v\rangle R^{2}}\xspace}

\newcommand{\unit}[1]{\ensuremath{\mathrm{\,#1}}\xspace}
\newcommand{\TeV}{\unit{TeV}}

\newcommand{\MeV}{\unit{MeV}}

\newcommand{\cm}{\unit{cm}}

\newcommand{\second}{\unit{s}}

\newcommand{\ph}{\unit{ph}}

\newcommand{\Figref}[1]{Figure \ref{figs:#1}}

\newcommand{\Eqnref}[1]{Eq.~(\ref{eqn:#1})}

\newcommand{\secref}[1]{section \ref{sec:#1}}

\title{Constraints on the pMSSM from LAT Observations of Dwarf Spheroidal Galaxies}
\author[a]{R.C. Cotta,}
\author[b]{A. Drlica-Wagner,}
\author[b]{\\S. Murgia,}
\author[b]{E.D. Bloom,}
\author[a]{J.L. Hewett,}
\author[a]{and T.G. Rizzo}
\affiliation[a]{Theory Group, SLAC National Accelerator Laboratory, 2575 Sand Hill Rd, Menlo Park, CA 94025, USA}
\affiliation[b]{W. W. Hansen Experimental Physics Laboratory, Kavli Institute for Particle Astrophysics and Cosmology, Department of Physics and SLAC National Accelerator Laboratory, Stanford University, Stanford, CA 94305, USA }
\emailAdd{randoo@slac.stanford.edu}
\emailAdd{kadrlica@stanford.edu}
\abstract{
We examine the ability for the Large Area Telescope (LAT) to constrain Minimal Supersymmetric Standard Model (MSSM) dark matter through a combined analysis of Milky Way dwarf spheroidal galaxies. We examine the Lightest Supersymmetric Particles (LSPs) for a set of $\sim 71$k experimentally valid supersymmetric models derived from the phenomenological-MSSM (pMSSM). We find that none of these models can be excluded at 95\% confidence by the current analysis; nevertheless, many lie within the predicted reach of future LAT analyses. With two years of data, we find that the LAT is currently most sensitive to light LSPs ($m_\mathrm{LSP}<50\gev$) annihilating into $\tau$-pairs and heavier LSPs annihilating into \bbbar.  Additionally, we find that future LAT analyses will be able to probe some LSPs that form a sub-dominant component of dark matter. We directly compare the LAT results to direct detection experiments and show the complementarity of these search methods.
}
\arxivnumber{}

\begin{document} 
\rightline{\vbox{\halign{&#\hfil\cr
&SLAC-PUB-14705\cr
}}}

\maketitle
\flushbottom
  \section{Introduction}
  \label{sec:intro}

Astrophysical evidence suggesting that non-baryonic dark matter (DM) comprises nearly 25\% of the energy density of the Universe is one of the most compelling arguments for particle physics beyond the Standard Model (SM)~\cite{silk}. At present, experimental tests of this DM component are almost exclusively limited to gravitational interactions, and few constraints exist on the character of DM. Axions, dark photons, sterile neutrinos and even more exotic theoretical constructs are all plausible DM candidates~\cite{Bergstrom:2010zz}, though models containing a new neutral and stable weakly-interacting massive particle (WIMP) of mass $\sim100\gev$ are by far the most studied. WIMPs are a favorable candidate because their mass and couplings to the SM can naturally give a cosmological relic density in agreement with the experimentally measured value~\cite{Komatsu:2008hk}. Additionally, WIMPs point to new physics at the weak scale ($\sim100\gev-1\,\tev$), a scale that has been the focus of much theoretical work to explain the stability of the Higgs potential and the origin of electroweak symmetry breaking.

In the past several decades, supersymmetry (SUSY) has been the most widely-studied, and arguably the best-motivated, theoretical framework for physics beyond the SM \cite{Haber:1984rc, Martin:1997ns,Chung:2003fi,Pape:2006ar,Dreiner:2008tw,Drees:2004jm,Baer:2006rs}. In the most attractive SUSY models, an extra matter parity (``R-parity") symmetry is used to simultaneously explain the stability of the proton and of the Lightest Supersymmetric Particle (LSP). In viable SUSY models the LSP is often the lightest neutralino ($\LSP$), which is one of the most widely studied examples of WIMP DM. Generic predictions of SUSY are difficult to obtain, since the minimal consistent SUSY extension of the SM, the Minimal Supersymmetric Standard Model (MSSM) introduces more than $100$ free parameters. A typical strategy for overcoming this difficulty is to highly constrain this set of parameters by employing aesthetic assumptions about the physical origin of SUSY at a very high energy (\ie, mSUGRA~\cite{Chamseddine:1982jx,Cremmer:1982vy}). In contrast, here we study a broader and more comprehensive subset of the MSSM, the phenomenological-MSSM (pMSSM) \cite{Berger:2008cq}. The pMSSM is derived from the MSSM using experimental data to eliminate parameters that are free in principle, but highly constrained by observations (\eg, sources of flavor violation in the new physics flavor sector). Thus, the pMSSM provides a compromise between the need to remain flexible and somewhat agnostic in assumptions about yet-undiscovered physics and the need to categorize the range of predictions made by well-motivated models. The LSPs of the pMSSM are viable candidates to comprise some or all of DM, and they may be probed through a variety of experimental approaches.

The possibility of DM-SM interactions having weak-force strength allows an exciting opportunity to detect and characterize the nature of DM via a combination of experimental efforts. For example, weak-strength interactions might lend themselves to study at the LHC, where DM particles could be produced and studied indirectly through missing energy signatures. Additionally, the DM halo permeating our galaxy could be detected directly through scattering interactions between DM particles and nuclei in detectors on Earth. Finally, indirect detection of DM is possible through the astrophysical observation of anomalous energetic SM particles resulting from DM particle annihilation (or decay). 

One of the most sensitive instruments for the indirect detection of DM is the Large Area Telescope (LAT) on board the \textit{Fermi Gamma-ray Space Telescope} (\Fermi). Gamma rays from the final state of DM annihilation (or decay) would be produced preferentially in regions of high DM density and may be detectable by the LAT. Dwarf spheroidal satellite galaxies (dSphs) of the Milky Way are promising targets for the detection of such a signal. These dSphs are DM-dominated and lack active astrophysical production of $\gamma$-rays~\cite{Mateo:1998wg,Grcevich:2009gt}, a troublesome background in many other searches for DM annihilation. The LAT Collaboration recently presented results constraining the annihilation cross section for a small set of prototypical DM models from a joint likelihood analysis of 10 dwarf spheroidal galaxies~\cite{dwarfs}. In the present paper, we extend this analysis to an investigation of $\sim71$k pMSSM models previously discussed in the literature~\cite{Berger:2008cq}.

We begin by briefly discussing the techniques employed to generate $\sim71$k pMSSM models and the various constraints imposed in their selection. We next describe the combined likelihood procedure for setting upper limits on the annihilation cross section for each pMSSM DM model using LAT observations of ten Milky Way dSphs. We compare the LAT cross section limits to the actual cross section for each pMSSM model and study the SUSY model dependence of these results in detail. The main findings are: (\emph{i}) that the LAT is currently most sensitive to light LSPs ($m_{\LSP}<50\gev$) annihilating primarily to $\tau$-pairs, (\emph{ii}) that annihilations to $\tau$-pairs are actually \emph{harder} to limit than annihilations to the other channels for relatively heavy ($m_{\LSP}>50\gev$) LSPs, and (\emph{iii}) that, surprisingly, a significant fraction of the models that are near LAT sensitivity have LSPs that would form a sub-dominant component of the total DM halo. Additionally, we discuss the relationship between LSP eigenstate composition and the LAT sensitivity. Finally, we compare expectations for near-future LAT dSph searches and direct detection experiments.

\section{Generation of the pMSSM Model Set}
\label{sec:pMSSM}

The sensitivity of the LAT detector can be explored over a broad region of supersymmetric parameter space. We investigate the $\gamma$-ray production from $\sim71$k points in the 19-dimensional pMSSM parameter space generated in previous work~\cite{Berger:2008cq}. These points pass all of the constraints discussed in this section and are referred to as pMSSM models.

The 19-dimensional parameter space of the pMSSM results from imposing the following minimal set of assumptions on the general R-Parity conserving MSSM \cite{Djouadi:2002ze}: ($i$) the soft parameters are taken to be real, allowing no new CP-violating sources beyond those in the CKM matrix; ($ii$) Minimal Flavor Violation \cite{D'Ambrosio:2002ex} is taken to be valid at the TeV scale; ($iii$) the first two generations of sfermions having the same quantum numbers are taken to be degenerate and to have negligible Yukawa couplings; and ($iv$) the LSP is taken to be the lightest neutralino and is assumed to be a stable thermal WIMP. No assumptions about the physics at high energy scales or SUSY-breaking mechanisms are employed. The first three conditions are applied to avoid issues associated with constraints from the flavor sector.  These assumptions reduce the SUSY parameter space to 19 free soft-breaking parameters that are given by the three gaugino masses, $M_{1,2,3}$, ten sfermion masses $m_{\tilde Q_1,\tilde Q_3,\tilde u_1,\tilde d_1,\tilde u_3,\tilde d_3,\tilde L_1,\tilde L_3,\tilde e_1,\tilde e_3}$, the three $A$-terms associated with the third generation  ($A_{b,t,\tau}$), and the usual Higgs sector parameters $\mu$, $M_A$ and $\tan \beta$.

The set of models discussed in this paper was selected by numerical scans over the 19-dimensional parameter space of the pMSSM. This selection required a choice of parameter range intervals and scan priors. Issues involved in this selection have been described in detail previously~\cite{Berger:2008cq,Cotta:2010ej,Conley:2010du}. Here, we simply note that two scans were performed: one employed a flat prior beginning with $10^7$ points, and a second used a logarithmic prior employing $2\times 10^6$ points. The relevant differences between these two scans are that ($i$)  all SUSY mass parameters were restricted to be $\leq 1$ TeV for the flat-prior case, while for the log-prior case this restriction was raised to $\leq 3$ TeV, and ($ii$) the choice of the logarithmic prior generally leads to more compressed sparticle spectra than does the flat-prior case.  Note that the restriction on the upper limit for the mass parameters in both scans is chosen to ensure relatively large production cross sections at the LHC. In the present work, we focus primarily on results for pMSSM models in the flat-prior set.

After scanning the 19-dimensional parameter space, we subjected the resulting points to a set of theoretical and experimental constraints to select models that are valid for study.  We briefly review these restrictions here.{\footnote {For full details, see ref.~\cite{Berger:2008cq}}} ($i$) Our theoretical constraints required that spectra must be tachyon free, color and charge breaking minima must be avoided, and a bounded Higgs potential must exist (leading to radiative electroweak symmetry breaking). ($ii$) We employed a number of constraints from the flavor sector and precision electroweak data arising from the measurements of $(g-2)_\mu$, $b\to s\gamma$, $B\to \tau \nu$, $B_S \to \mu^+\mu^-$, meson--anti-meson mixing, the invisible width of the $Z$ and $\Delta \rho$. ($iii$) Restrictions resulting from numerous direct searches at LEP for both the SUSY particles themselves, as well as the extended SUSY-Higgs sector, were imposed. Some of these searches needed to be re-evaluated in detail to remove SUSY model-dependent assumptions~\cite{Berger:2008cq}. ($iv$) Null results from the set of Tevatron SUSY sparticle and Higgs searches were imposed. The most restrictive Tevatron data came from searches for stable charged particles~\cite{Abazov:2008qu} and from searches for an excess of multijet events with missing transverse energy~\cite{Abazov:2007ww}. We note that in the latter case, the search strategies were designed for kinematics expected in mSUGRA-inspired models, which required specialized simulations to apply these results in the context of the pMSSM~\cite{Berger:2008cq}. While these bounds have been superseded by LHC analyses, we note that collider searches have been found to be largely uncorrelated with the most important DM observables (relic density, annihilation and scattering cross-sections)~\cite{Cotta:2011ht}, and we expect the results presented here to accurately represent the prospects for indirect detection in the context of the pMSSM. ($v$) Finally, we have required that the LSP contribution to the dark matter relic density not exceed the upper bound determined by WMAP~\cite{Komatsu:2008hk} and that the LSP scattering cross sections obey concurrent constraints from direct detection experiments.  

\begin{figure}[t]
  \centering
  \includegraphics[width=0.6\textwidth]{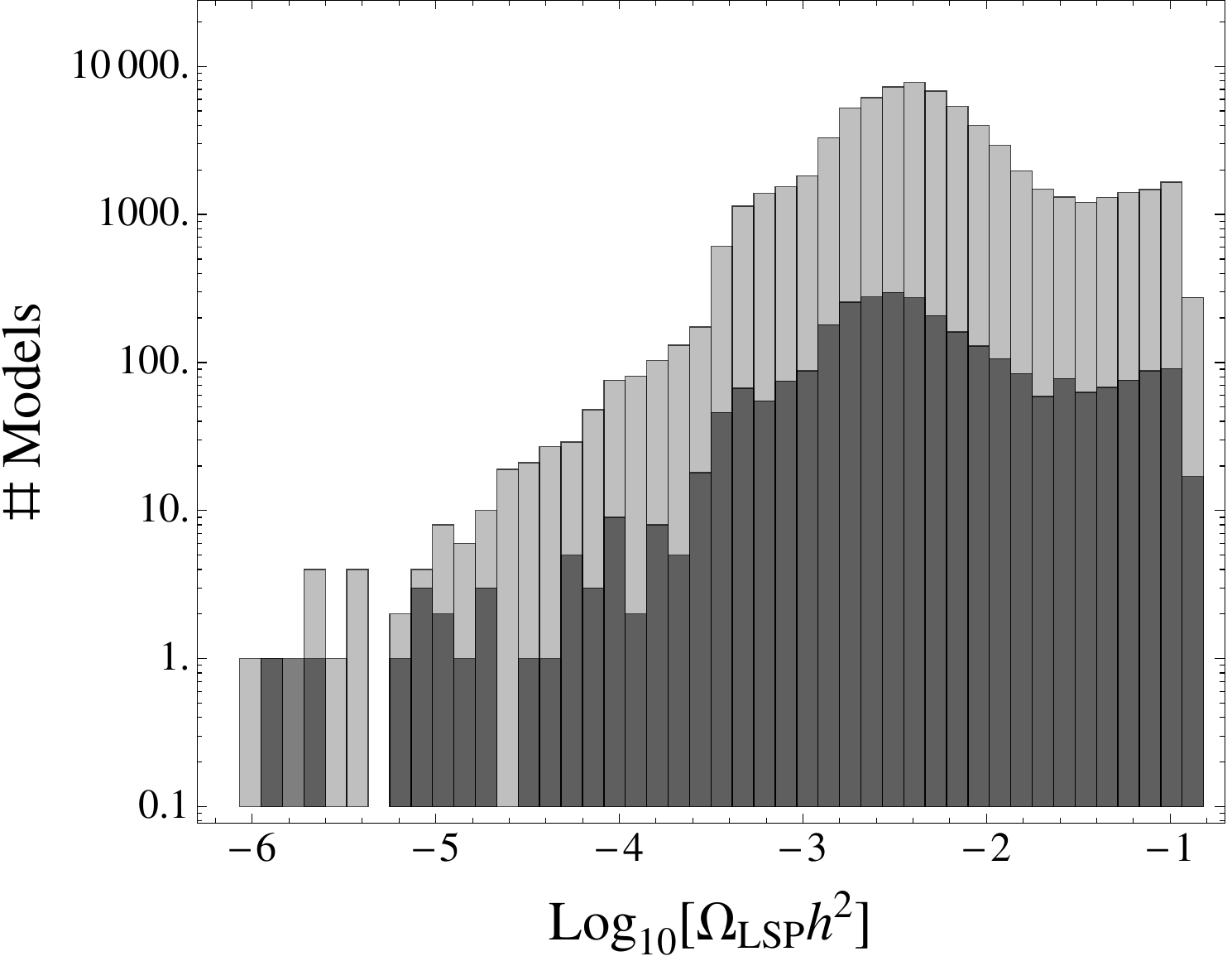}
  \caption{Logarithm of the LSP relic density, $\Omega_\mathrm{LSP} h^2$, for the flat-prior selected pMSSM models (grey bars) and log-prior selected pMSSM models (black bars).}
  \label{figs:rd}
\end{figure}

After imposing theoretical and experimental constraints, $\sim 68.4$k models from the flat-prior sample and $\sim 2.9$k models from the log-prior sample remain for study. Note that the LSPs of the pMSSM models are not required to saturate the measured relic density. Thus, many models allow for multicomponent DM, a prosaic example being composed in part by a pMSSM neutralino and in part by an axion that solves the strong CP problem. The distributions of LSP relic densities found in our flat- and log-prior model sets are shown in \Figref{rd}. We find that a small subset of pMSSM models saturate the WMAP bound (with $\Omega_\mathrm{LSP} h^2 > 0.1$).

Gamma-ray energy spectra are calculated for all $\sim 71$k pMSSM models using the computational package DarkSUSY 5.0.5 \cite{Gondolo:2004sc}. DarkSUSY calculates the total $\gamma$-ray yield from annihilation, as well as the rates into each of 27 final state channels. We omit contributions from the loop-suppressed monochromatic channels $\gamma\gamma$ and $\gamma Z^0$ which, although distinctive, are typically tiny in our model set. These $\gamma$-ray spectra are tested for consistency with the LAT $\gamma$-ray data.

  \section{LAT  $\gamma$-ray Limits on the pMSSM}
  \label{sec:fermi}

The Milky Way dSphs are a promising set of sources for the indirect detection of DM via $\gamma$ rays. Stellar velocity data from these galaxies suggest that they are very rich in DM, while observations at other wavelengths show no signs of astrophysical signals~\cite{Mateo:1998wg,Grcevich:2009gt}. The integrated signal flux at the LAT, $\phi_s$ ($\ph\,\cm^{-2}\,\second^{-1}$), from pair annihilation in a DM distribution with density given by $\rho(\vec{r})$ is
  \begin{equation}
   \phi_s =
    \underbrace{\int_{\Delta\Omega}\Big\{\int_{\rm l.o.s}\rho^{2}(\vec{r})dl\Big\}d\Omega '}_{J}\cdot\frac{1}{4\pi}\frac{\sigv_T}{2m_{\LSP}^{2}}\int^{E_{max}}_{E_{min}}\frac{dN^{\gamma,tot}}{dE_{\gamma}}dE_{\gamma},
    \label{eqn:gamrate}
    \end{equation}
The preceding J-factor represents the line-of-sight integral through $\rho^2$ integrated over a solid angle $\Delta\Omega$, while the second factor is strictly dependent on the particle physics properties of the DM model and is integrated over the experimental energy threshold (discussed in more detail in \secref{susymd}). Using \Eqnref{gamrate}, it is possible to combine the integrated J-factor for a dark matter distribution (calculated empirically for the dSph) with the particle physics characteristics of a DM model (from the pMSSM model set) to describe the predicted $\gamma$-ray annihilation signal.

We follow the procedure of Ackermann et al. (2011; henceforth A11) to constrain the $\gamma$-ray signal from ten dSphs with a joint likelihood analysis of the LAT data.  Our data sample and event selection are identical to those described in A11, taking photons in the energy range from $200\,\MeV < E < 100\gev$. In concordance with A11, we use LAT ScienceTools\footnote{http://fermi.gsfc.nasa.gov/ssc/data/analysis/software} version v9r20p0 and the P6\_V3\_DIFFUSE IRFs.\footnote{http://fermi.gsfc.nasa.gov/ssc/data/analysis/scitools/overview.html} J-factors and associated uncertainties for the ten dSphs are taken from Table 1 of A11, where they were calculated using line-of-sight stellar velocities and the Jeans equation~\cite{dwarfs}.

Our procedure for constraining \sigv  differs from that of A11 in that we model the $\gamma$-ray emission from the dSphs with spectra generated from the $\sim 71$k pMSSM models rather than prototypical annihilation channels (\ie, \bbbar, $\tau^{+}\tau^{-}$ etc.). We calculate a joint likelihood for each pMSSM model by tying the pMSSM model parameters across the regions of interest (ROIs) surrounding the ten dSphs.  Following A11, we incorporated uncertainties in the J-factors of the dSphs as nuisance parameters in our likelihood maximization.  Thus, our joint likelihood function is
\begin{equation}
\label{eqn:likelihood}
\begin{aligned}
  L(D\,|\,{\bf p_m},\{{\bf p_k} \}) = \prod_k &L_k^{\rm LAT}(D_k\,|\,{\bf p_m},{\bf p_k})  \\ 
                                              &\times \frac{1}{\ln(10) J_k \sqrt{2 \pi} \sigma_k} e^{-(\log_{10}(J_k)-\overline{\log_{10}(J_k)})^2/2\sigma_k^2} . 
\end{aligned}
\end{equation}
Here, $k$ indexes the ROIs, $L^{\rm LAT}_k$ denotes the standard LAT binned Poisson likelihood for the analysis of a single ROI,\footnote{http://fermi.gsfc.nasa.gov/ssc/data/analysis/documentation/Cicerone/Cicerone\_Likelihood/} $D_k$ represents the binned $\gamma$-ray data, \{$\bf{p_m}$\} represents the set of ROI-independent pMSSM model parameters, and \{$\bf{p_k}$\} are the ROI-dependent model parameters. Included in \{$\bf{p_k}$\} are both the flux normalizations of background $\gamma$-ray sources (diffuse and point-like) and the associated dSph J-factors and uncertainties. We find no significant $\gamma$-ray signal from any of the dSphs when analyzed individually or jointly for any of the pMSSM models.

For each of the $\sim 71$k pMSSM models, we calculate the maximum annihilation cross section, \sigvm, consistent with the null detection in the composite LAT data. To deal with nuisance parameters present in the joint likelihood, we constructed a profile likelihood $L_p(D\,|\,\sigv)$ by scanning in \sigv and maximizing $L(D\,|\,{\bf p_m},\{{\bf p_i}\})$ with respect to the other free parameters~\cite{Rolke:2004mj}. For each model, we proceeded to obtain a 95\% one-sided confidence interval on the value of \sigv by first maximizing $\log(L_p)$ with respect to \sigv and then evaluating $\log(L_p)$ at increasing values of \sigv until $\Delta \ln L_p = -2.71/2$~\cite{Rolke:2004mj}. This one-sided 95\% confidence limit on \sigv serves as our value of \sigvm, which is compared to the true value annihilation cross section for each pMSSM model.

  \section{Results}
  \label{sec:results}

In this section, we investigate the ability to constrain the predicted flux spectra of pMSSM models using LAT $\gamma$-ray flux measurements from ten dSphs. We then discuss the SUSY model dependence of our results and improvements that can be made in searches such as this one to enhance sensitivity to annihilations from certain classes of models. 
   
\subsection{Model Constraint Distance}
\label{sec:basic} 
 
We present limits placed on each of the $\sim 71$k pMSSM model spectra from a joint likelihood fit of ten Milky Way dSphs as described in Section \ref{sec:fermi}. For each model, we compute the effective distance from constraint as the ratio 
\begin{equation}
\bigd\equiv\sigvm/\sigv_T,
     \label{eqn:bigddef}
\end{equation}
 which compares the maximum annihilation cross section allowed by the data for a given spectral shape, $\sigvm$, to the actual cross section predicted for the pMSSM model giving rise to this spectral shape, $\sigv_T$. Here, the symbol $\sigv_T$ represents the current annihilation cross section obtained assuming that the LSP follows a standard thermal cosmological evolution.\footnote{One should not confuse the ``$T$" in $\sigv_T$, here denoting thermal cosmology, for the thermal average over the current halo Boltzmann distribution, which is here denoted by the angle brackets $\langle\rm{\ldots}\rangle$.} It is important to note that, because we have employed the WMAP measurement of the DM relic density only as an upper bound when selecting pMSSM models, the appropriate normalization of the $\gamma$-ray spectra for each model involves a rescaling of the empirical estimate for the local DM energy density by the factor
\begin{equation}
 R=\frac{\Omega_{\chi}}{\Omega_{\rm{WMAP}}},~~~\mathrm{as,}~~~\sigv_T\equiv\tsigv,
     \label{eqn:OmegaR}
\end{equation}
where $\Omega_{\rm{WMAP}}h^2=0.1143$ \cite{Komatsu:2008hk}. The factors of $R$ in $\sigv_T$ serve to obtain the correct thermal number density of WIMPs $n=\rho_{\LSP}/m_{\LSP}=(\rho_0 R) /m_{\LSP}$ (where $\rho_{\LSP}$ and $\rho_0\approx0.3\gev\cm^{-3}$ are the energy density of the LSP and the empirically determined \cite{Gates:1995dw} total local DM energy density, respectively). LSPs in models with $R \ll 1$ are interpreted as comprising one component of a multicomponent DM halo. For definiteness, we identify pMSSM models with LSPs that nearly saturate the WMAP relic density, $\Omega_{\rm{LSP}}h^2>0.10$ ($R>0.875$), and distinguish this subset of models in the figures that follow. Deviations from the scaling in \Eqnref{OmegaR} would be appropriate for models where non-standard cosmology or other non-standard mechanisms~\cite{Feldman:2008xs,Ibe:2008ye,Guo:2009aj,Kadota:2010xm,Grajek:2008pg,Kane:2009if,Acharya:2009zt} sever the relationship between the relic density and the present-day annihilation cross section obtained from thermal Boltzmann cosmological evolution. We assume that standard thermal evolution occurs and calculate results relative to this scenario.

\begin{figure}[t]
    \centering
    \includegraphics[width=1.0\textwidth]{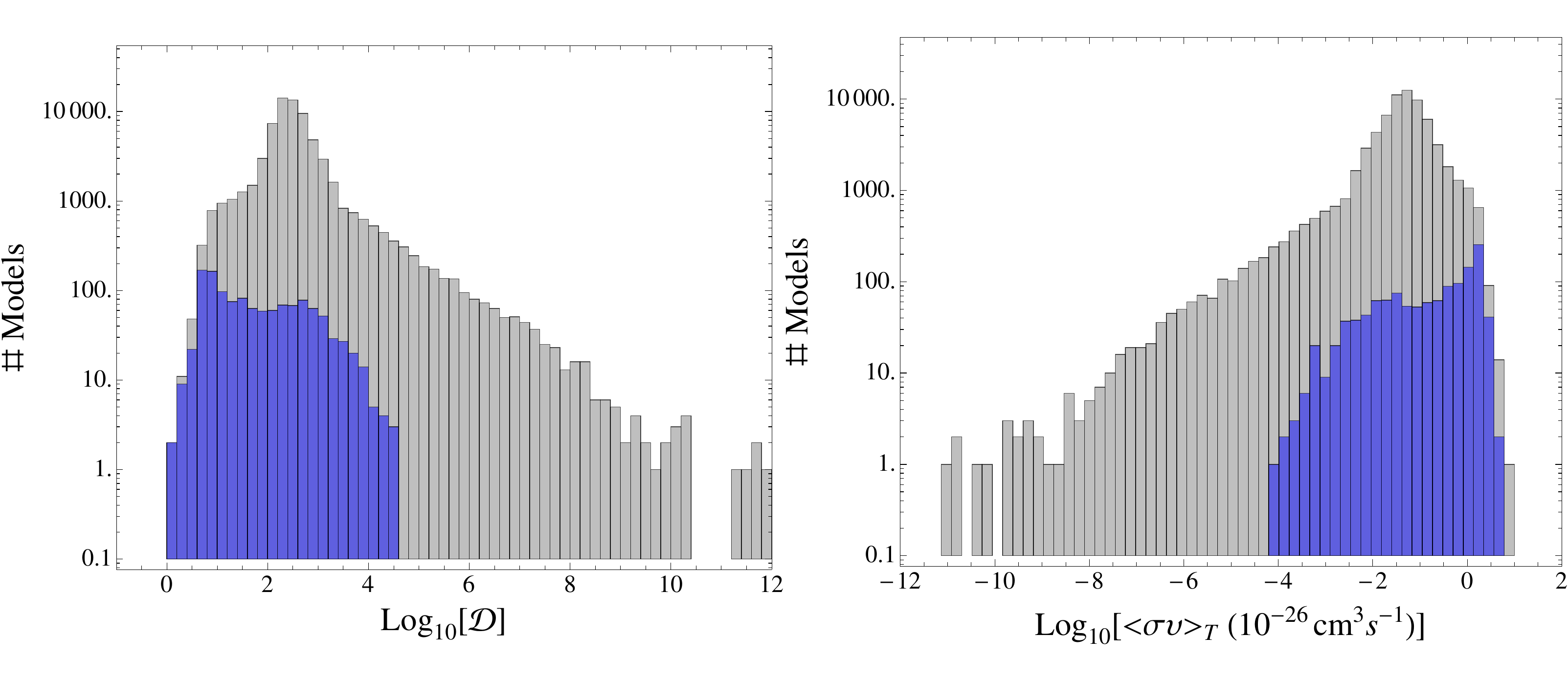}
    \caption{Distance from constraint, $\bigd$ (\Eqnref{bigddef}), for all flat-prior selected pMSSM models (grey bars) and flat-prior models with $R\approx1$ (blue bars). We similarly histogram $\sigv_T$ and note that the $\bigd$ distribution is largely driven by the $\sigv_T$ distribution.}
    \label{figs:basichisto}
  \end{figure}

The ratio $\bigd$ (\Eqnref{bigddef}) is calculated for each pMSSM model and is displayed in \Figref{basichisto}. Models with $\bigd<1$ would be excluded at $> 95\%$ confidence, while models with $\bigd>1$ evade this limit by a factor of $\bigd$. None of the pMSSM models are excluded at 95\% confidence by this analysis; however, values of $\bigd$ reach $\sim1.5$ for many of the models with LSP masses $\lsim40\gev$, relic densities nearly saturating the WMAP measurement ($R\approx1$) and annihilating predominantly to $\tau$-pairs (as is further discussed in \secref{susymd}).

Since the LAT is presently very close to constraining part of the pMSSM parameter space, it is useful to estimate how constraints may improve over a 10 year mission lifetime. In the low-energy, background dominated regime, the LAT point source sensitivity increases as roughly the square-root of the integration time. However, in the high-energy, limited background regime (where many pMSSM models contribute), the LAT sensitivity increases more linearly with integration time. Thus, 10 years of data could provide a factor of $\sqrt{5}$ to 5 increase in sensitivity. Additionally, optical surveys such as Pan-STARRS and the Dark Energy Survey could provide a factor of 3 increase in the number of Milky Way dSphs corresponding to an increased constraining power of $\sqrt{3}$ to 3~\cite{Tollerud:2008ze}.  Ongoing improvements in LAT event reconstruction, a better understanding of background contamination, and an increased energy range are all expected to provide additional increases in the LAT sensitivity. Thus, we find it plausible that the LAT constraints could improve by a factor of 10 compared to current constraints, and we choose to examine pMSSM models with $\bigd < 10$ in detail. The interest in models with $\bigd < 10$ is additionally motivated by predictions that DM substructure may increase the J-factors of dSphs by a factor of 2 to 10~\cite{Anderson:2010df,Martinez:2009jh}. Such a boost would translate into a tightening of the current upper limits on \sigv by a comparable factor. 

In Figure \ref{figs:basichisto} we observe, as expected, that the relic density of a given pMSSM model is important in determining the $\gamma$-ray signal strength. Despite this, we see that that the range of predictions in either subset of models is quite large, spanning many orders of magnitude. We note the surprising fact that, among models with $\bigd<10$, nearly 40\% do not saturate the WMAP bound.

When we compare our set of log-prior models to the flat-prior set, we find that the results are generally consistent with a somewhat larger tail towards large values of $\bigd$. As noted in the next section, the shape of the $\bigd$ distribution is largely determined by the shape of the $\sigv_T$ distribution, and one can verify that the difference in $\bigd$ distributions between flat- and log-prior cases is echoed in their $\sigv_T$ distributions. Models with such large $\bigd$ values are seen to be special cases, involving a finely tuned relationship between the DM mass and the mass of one or more of the SUSY Higgs states (such that $m_{\LSP}\sim m_h/2$, so-called``Higgs funnel'' models), that annihilate so efficiently in the early universe as to have minuscule relic density.

\subsection{SUSY Model Dependence}
\label{sec:susymd} 

\begin{figure}[t]
    \centering
    \includegraphics[width=0.50\textwidth]{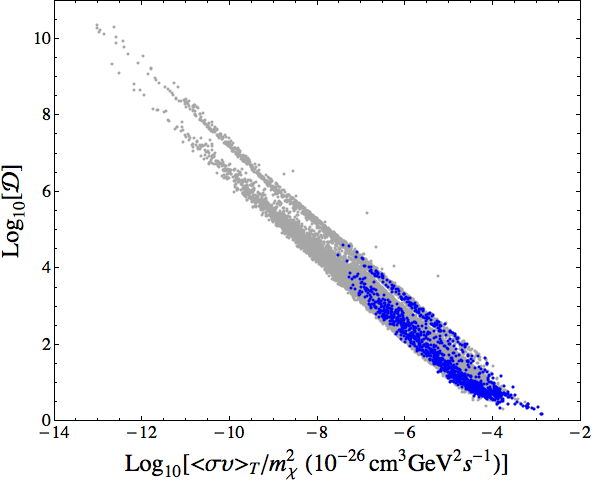}
    \caption{
Flat-prior pMSSM models represented in the $\bigd$ vs.\ $\sigv_T/2m_{\LSP}^{2}$ plane. Grey points represent generic models in this set while the subset of models with $R\approx 1$ are highlighted in blue. One observes that the wide range of $\bigd$ values corresponds directly to the wide range $\sigv_T/2m_{\LSP}^{2}$ values in our model set, and that at a given value of $\sigv_T/2m_{\LSP}^{2}$ there is about an order of magnitude span in $\bigd$ values. It can also be noted that only $\sim 60\%$ of models with $\bigd < 10$ have $R\approx 1$.}
    \label{figs:ratiovssignal} 
  \end{figure}

In discussing the SUSY model dependence of these results, we first note (\cf, \Figref{ratiovssignal}) that most of the span in $\bigd$ arises from the wide ranging values of the quantity $\sigv_T/2m_{\LSP}^{2}$. For a given value of $\sigv_T/2m_{\LSP}^{2}$ we observe that there is only about an order of magnitude span in $\bigd$. To understand this finding we note that the total signal $\gamma$-ray flux, $\phis$, as calculated in \Eqnref{gamrate} (and repeated here for clarity), is
  \begin{equation}
   \phi_s =
    \underbrace{\int_{\Delta\Omega}\Big\{\int_{\rm l.o.s}\rho^{2}(\vec{r})dl\Big\}d\Omega '}_{J}\cdot\frac{1}{4\pi}\frac{\sigv_T}{2m_{\LSP}^{2}}\int^{E_{max}}_{E_{min}}\frac{dN^{\gamma,tot}}{dE_{\gamma}}dE_{\gamma},
    \end{equation}
We focus now on the particle physics-dependent piece, which is itself a product of the factor $\sigv_T/2m_{\LSP}^{2}$ and of the integral over the the total $\gamma$-ray continuum yield curve:\footnote{
Here the sum is over annihilation final-state channels with terms describing
  hadronization yield (``secondary'' $\gamma$ rays), final-state radiation (FSR) and virtual internal
  bremsstrahlung (VIB). As previously noted, a possible monochromatic $\gamma$ contribution is negligible here. We use the language of \cite{Bringmann:2007nk} in discriminating FSR and VIB, although it has
    been pointed out that such a distinction is somewhat artificial (or not even gauge invariant)
    \cite{Barger:2009xe}.}
  \begin{equation}
    \frac{dN^{\gamma,tot}}{dE_{\gamma}} =
    \sum_{i}B_{i}\bigg\{\frac{dN_{i}^{\gamma,sec}}{dE_{\gamma}} +
    \frac{dN_{i}^{\gamma,FSR}}{dE_{\gamma}} + \frac{dN_{i}^{\gamma,VIB}}{dE_{\gamma}}\bigg\}.
    \label{totyield}
  \end{equation}
 The $\bigd$ value expected from a given model is thus \emph{predominantly determined} by the total thermally-averaged annihilation cross section, $\sigv_T$, the relic density (via $R$) and the LSP mass (via $\sigv_T/2m_{\LSP}^{2}$), and \emph{to a much lesser extent} by the shape of the SUSY signal spectrum.

We next discuss the SUSY model dependence of the spectral \emph{shape} of the annihilation signal. It is useful to remove the large prefactor, $\sigv_T/m^2_{\LSP}$, focusing on the quantity: 
  \begin{equation}
    \rphis\propto
\int^{E_{max}}_{E_{min}}\frac{dN^{\gamma,tot}}{dE_{\gamma}}dE_{\gamma}\;,
    \label{eqn:theintegral}
    \end{equation}
where we integrate the total continuum yield curve over the
 experimental energy range from $E_{min}=200\mev$ to $E_{max}=100\gev$. In translating to $\rphis$, we have removed the explicit dependence on the total annihilation rate $\sigv_T$ and most of the dependence on the LSP mass. What remains is a quantity that depends on the relative annihilation rates into SM final state channels, which is useful for comparing models with similar $\sigv_T$ and LSP mass. A small residual LSP mass dependence remains in $\rphis$, via the relationship between the LSP mass and the limits of integration in \Eqnref{theintegral}. 
 
The supersymmetric origin of our $\gamma$-ray spectra is reflected in the \emph{distribution} of annihilations into distinct SM final states ($f$). In SUSY, due to the Majorana nature of the annihilating $\LSP$ particles and the fact that $\LSP$s are non-relativistic in current DM halos, the annihilation rates for processes such as $\LSP\LSP\to f\bar{f}$ are proportional to $(m_f/m_{\LSP})^2$, a fact that is often referred to as helicity suppression \cite{Goldberg:1983nd}. The ratio of rates into distinct channels (say $f\bar{f}$ and $f'\bar{f'}$) is thus $\propto(m_f/m_{f'})^2$ so that annihilation rates into channels with heavy final-state SM particles often dominate those with lighter final-state SM particles. These ratios are complicated by the fact that the rates $\LSP\LSP\to f\bar{f}$ are coherent sums of subprocesses with varying couplings, mixing angles, and exchanged particle masses. Nonetheless, in our pMSSM model set, we see that the dominant annihilation final states are often in accord with the helicity suppression intuition. For the vast majority of models in our set, annihilations are predominantly composed of a mixture of the $b\bar{b}$, $\tau^+\tau^-$, $W^+W^-$, $Z^0Z^0$ and $t\bar{t}$ channels that are kinematically allowed (\ie, $m_{\LSP} > m_{f}$). We observe a number of cases where the loop-level annihilation to gluons, $\LSP\LSP\to gg$, is dominant\footnote{Loop-level annihilation to glouns can occur when the LSP co-annihilates with light flavored sfermions (\eg, $\tilde{u}$, $\tilde{d}$, $\tilde{e}$, $\tilde{\nu}$, etc.) in the early Universe. In current DM halos, annihilation by exchange of the lightest sfermions is heavily helicity suppressed ($\sim m^2_f/m^2_{\LSP}$) so that the loop-level process $\LSP\LSP\to gg$ may become the most efficient annihilation channel today.} and a small number of cases where there are sizable (but still sub-dominant) contributions from the $hA^0$, $HA^0$, $hZ^0$, $HZ^0$, $W^{\pm}H^{\mp}$ and (monochromatic) $\gamma\gamma$ or $\gamma Z^0$ channels. 

\begin{figure}[t]
    \centering
    \includegraphics[width=1.00\textwidth]{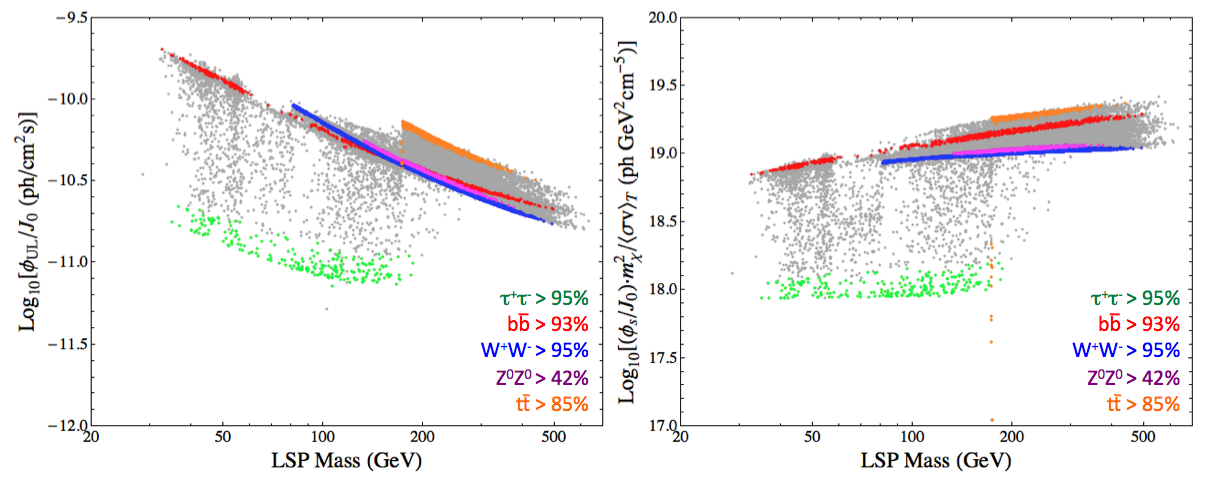}
    \caption{Flat-prior pMSSM models in the LAT flux upper limit vs.\ LSP mass (left panel) and scaled signal flux vs.\ LSP mass (right panel) planes (taking $J_0=10^{19}\gev^2\cm^{-5}$ as a reference J-factor). The full flat-prior model set is displayed as grey points and models whose annihilations occur predominantly through a given final state channel are overlaid in other colors. Models with $\langle\sigma v\rangle_{\tau\bar{\tau}}/\langle\sigma v\rangle>0.95$ (green), with $\langle\sigma v\rangle_{b\bar{b}}/\langle\sigma v\rangle>0.93$ (red), with $\langle\sigma v\rangle_{W^+W^-}/\langle\sigma v\rangle>0.95$ (blue), with $\langle\sigma v\rangle_{Z^0Z^0}/\langle\sigma v\rangle>0.42$ (magenta) and with $\langle\sigma v\rangle_{t\bar{t}}/\langle\sigma v\rangle>0.85$ (orange) are shown. Purities are chosen to obtain model subsets of similar size.}
    \label{figs:fluxesnlimits}
  \end{figure}
  
The ability of the LAT to constrain a given pMSSM model can be decomposed into two planes (displayed in \Figref{fluxesnlimits}). The $\phiul$ (LAT flux upper limit) vs.\ LSP mass plane demonstrates the ability of the LAT to constrain the spectral shapes of various annihilation channels, while the $\rphis$ vs.\ LSP mass plane describes the (scaled) signal flux expected from each channel. We see that subsets of models with nearly pure annihilation final state channels cluster tightly in both the $\phiul$ vs.\ LSP mass and $\rphis$ vs.\ LSP mass planes and, with the exception of the dominantly $Z^0Z^0$ channel models,\footnote{While the subset of models with $(\sigv_{ZZ}/\sigv)>0.42$ do not extend down to $m_{\LSP}\approx m_{Z}$, the subset of models with $(\sigv_{ZZ}/\sigv)>0.20$ do extend to this kinematic endpoint. The highest purity for $\LSP\LSP\to Z^0Z^0$ annihilations in our model set is $(\sigv_{ZZ}/\sigv)\approx0.45$, such models also annihilate to $\LSP\LSP\to W^+W^-$ with $(\sigv_{WW}/\sigv)\approx0.55$.} extend down to their kinematic endpoints.\footnote{The set of $\mathcal{O}(10)$ orange points near the top threshold, $m_{\LSP}\approx m_t$, are models that annihilate dominantly through the $\LSP\LSP\to t\bar{t}$ channel. They have very bino-like LSPs (supressing many other channels) and currently annihilate dominantly through stop exchange. These models satisfy the WMAP relic abundance constraint either by co-annihilation with a light stop or via the exchange of very light sfermions (\ie, channels that were more efficient at freeze-out). Since they are forced to annihilate dominantly to $t\bar{t}$ with $m_{\LSP}\approx m_t$, they are phase-space suppressed, $\langle\sigma v\rangle_{t\bar{t}}\propto(1-m^2_t/m^2_{\chi})^{1/2}$, and their fluxes are much lower than typical models that annihilate through this channel.}

The left panel of \Figref{fluxesnlimits} confirms that the LAT search places tighter constraints ($\phiul$) on harder $\gamma$-ray spectra, which is to be expected as astrophysical backgrounds fall rapidly with energy. In particular, we see that LAT constraints on spectra from nearly pure $\LSP\LSP\to \tau^+\tau^-$ annihilations are tighter, by about an order of magnitude, than those placed on nearly pure $\LSP\LSP\to b\bar{b}$ annihilations. One expects the shapes and the relationship between these two ``curves" to change significantly as LAT data taking continues. The LAT sensitivity to hard spectra, which contribute a significant number of photons in the background-free regime ($\gsim10\gev$), is expected to increase more quickly than the sensitivity to softer spectra, which contribute in the background-dominated regime. Thus, the gap between limits on $\LSP\LSP\to \tau^+\tau^-$ and $\LSP\LSP\to b\bar{b}$ in the left panel of \Figref{fluxesnlimits} is expected to widen with increased LAT data taking.

\begin{figure}[t]
   \centering
   \includegraphics[width=0.50\textwidth]{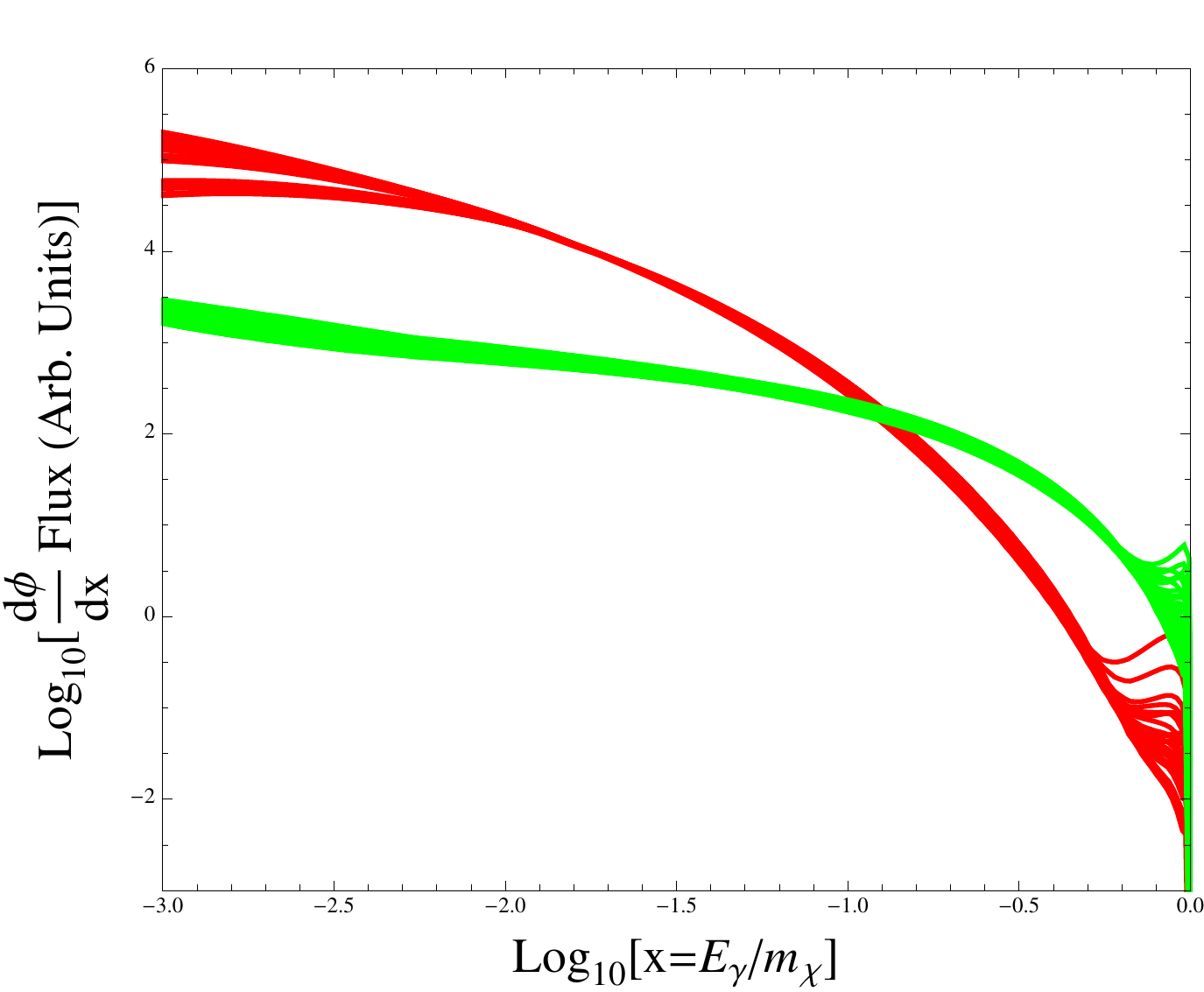}
   \caption{We display spectral shapes for pMSSM models that
     annihilate into $b\bar{b}$ (red) and $\tau$-pairs (green) final states with
     purities as in \Figref{fluxesnlimits}. Spectra have been scaled to
     remove the $\sigv_T/m_{\LSP}^2$ prefactor and plotted in terms of the
     variable $x=E_{\gamma}/m_{\LSP}$ in order to emphasize the
     universality of individual final state spectra.}
   \label{figs:shapes}
 \end{figure}

We note from the right panel of \Figref{fluxesnlimits} that nearly pure annihilations to $\tau$-pairs yield about an order of magnitude \emph{fewer} signal photons than models with nearly pure annihilations to $b\bar{b}$, at the same LSP mass and $\sigv_T$. This
 finding was discussed at length in ref.~\cite{Cotta:2010ej}, where
 it was noted that, although the $\gamma$-ray spectra resulting from
 $\tau$-like annihilations are harder (due to a large contribution
 from prompt $\pi^0$ decay), they are also shallower at low energies.
 This is demonstrated in \Figref{shapes}, where we display spectra
 (as calculated by DarkSUSY 5.0.4) for models that
 annihilate nearly purely into the $b\bar{b}$ and $\tau$-pair
 final-state channels. The curves displayed here correspond to the
 integrand in \Eqnref{theintegral} and, when plotted in terms of the
 variable $x\equiv E_{\gamma}/m_{\LSP}$, have a nearly universal shape
 (there is significant SUSY model dependence in the signal spectra
 for $E_{\gamma}\approx m_{\LSP}$, due to internal bremsstrahlung
 \cite{Bringmann:2007nk}). Comparing the $b\bar{b}$ and $\tau$-pair
 cases it is clear that, at a given LSP mass
 and $\sigv_T$, the softer $b\bar{b}$ spectra will produce a much
 larger integrated flux of signal $\gamma$ rays than the $\tau$-pair
 spectra. We also observe that the vast majority of signal $\gamma$
 rays have energy $E_{\gamma}\ll m_{\LSP}$, regardless of the
 particular final-state channel, so that the integrated signal flux
 depends sensitively on the lower limit of integration
 $x_{min}\equiv E_{min}/m_{\LSP}$. Heavier LSPs allow integration to lower $x$ values and thus a wider gap between the total flux from $b\bar{b}$ and
$\tau$-pair spectra. This is reflected in the relative slope of
 the red and green ``lines'' in the right panel of
 \Figref{fluxesnlimits}. 

\begin{figure}[t]
   \centering
   \includegraphics[width=1.0\textwidth]{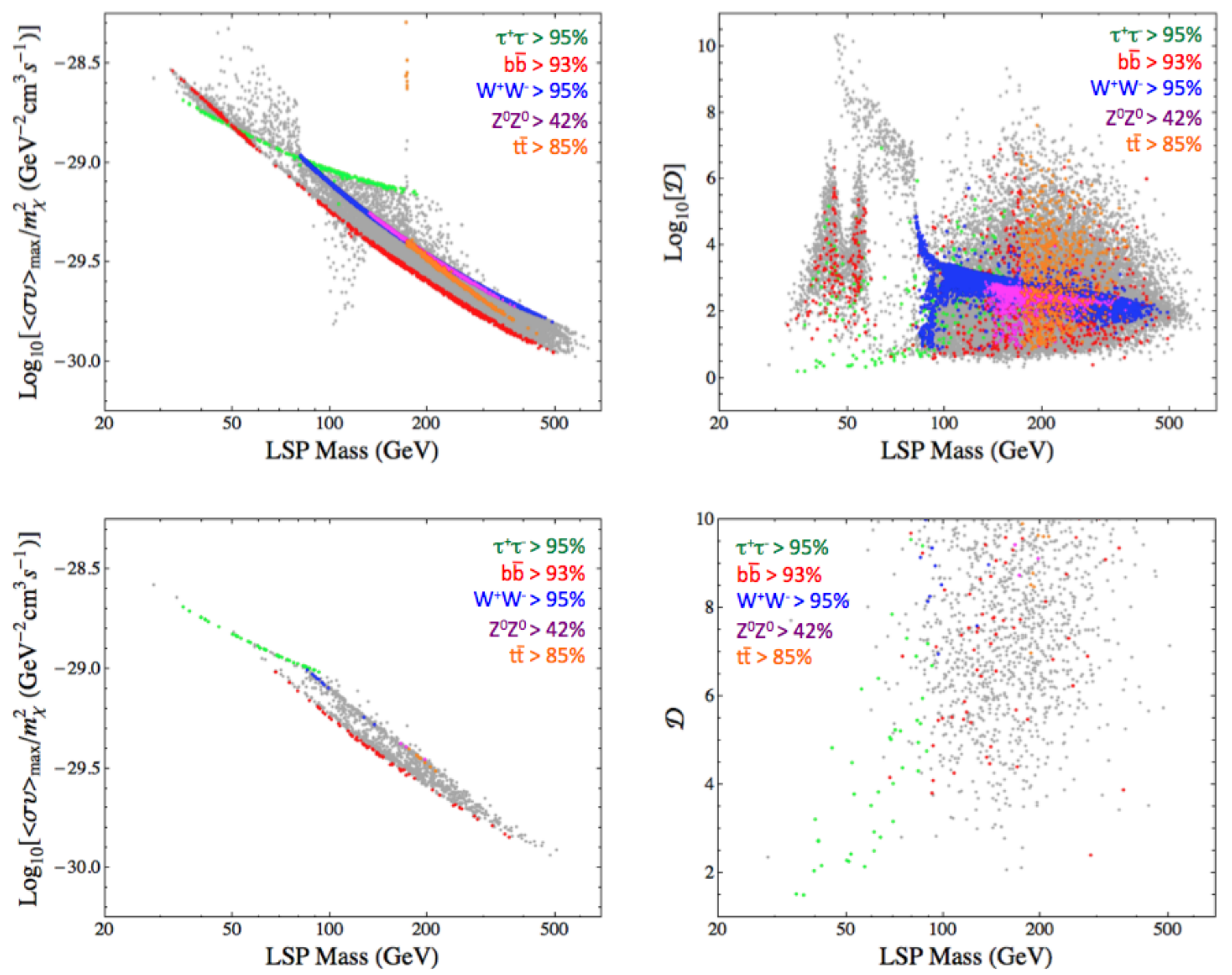}
   \caption{We display points representing pMSSM models in the $\sigvm/m^2_{\LSP}$ vs.\ LSP mass (left panels) and $\bigd$ vs.\ LSP mass (right panels) planes. The full flat-prior model set is displayed as grey points and models whose annihilations occur predominantly through a given final state channel are overlaid in other colors, as denoted in the figure. In the left panels, one can see that removing the dependence on total annihilation rate and LSP mass (scaling $\bigd\times\sigv_T/m^2_{\LSP}=\sigvm/m^2_{\LSP}$)  allows for tight localization of models with similar annihilation spectra, whereas it is comparatively difficult to predict where models fall in the $\bigd$ vs.\ LSP mass plane without such scaling. The upper panels display these relations for all pMSSM models while the lower panels zoom in on those models that are closest to constraint ($\bigd < 10$).}
   \label{figs:annihilations}
 \end{figure}

The two panels of \Figref{fluxesnlimits} describe two important
factors in determining $\bigd\equiv\sigvm/\sigv_T\equiv\phiul/\phis$:
(\emph{i}) the ability to tightly constrain the flux from a given
spectral shape ($\phiul$) and (\emph{ii}) the flux that can be expected
from that particular spectral shape ($\phis$). These pieces of information must be combined to determine the ease with which the LAT can constrain final state channels (\ie, the magnitude of $\bigd$), while providing a fair comparison between models annihilating to different final states with similar LSP mass and
$\sigv_T$. This is shown in the left panels of \Figref{annihilations}.

In the left panels of \Figref{annihilations} we display points for our models in the $\bigd\times\sigv_T/m^2_{\LSP}$ vs.\ LSP mass plane, with models that annihilate largely into single final state channels colored as in \Figref{fluxesnlimits}. Using $\bigd\times\sigv_T/m^2_{\LSP}$ is equivalent to using $\phiul/(\rphis)$, a ratio of the values found in either panel of \Figref{fluxesnlimits}. We see that annihilations to a given final state are organized nicely in this plane. For comparison, in the right panels of \Figref{annihilations} we display points for each of our flat-prior models in the $\bigd$ vs.\ LSP mass plane using the same color scheme,where the colored points are seen to be highly mixed. \Figref{annihilations} allows us to predict whether it is easier to constrain annihilations to $\tau$-pairs or to $b\bar{b}$ (for example), at a given LSP mass, $\sigv_T$, and with experimental thresholds  $E_{min}$ and $E_{max}$. From the left panels we observe that  annihilations to $\tau$-pairs are surprisingly \emph{more difficult} to constrain than annihilations to $b\bar{b}$ for LSP masses $m_{\LSP}\gsim 50\gev$, while the opposite is true for lighter LSPs. This crossover at $m_{\LSP}\approx 50\gev$ is the point at which there is a balance between the relative ease of constraining the $\tau$-like spectral shape and the relatively low number of signal $\gamma$-rays produced in these annihilations, relative to the $b\bar{b}$ case. With longer observations, we expect this crossover to move toward higher energies as the limits on harder spectra are expected to tighten more quickly than limits on softer spectra.

We emphasize that the grey points in the left panels of \Figref{annihilations} are as important as the colored bands. These regions represent the combination of annihilation final-state channels realized in the pMSSM model set and provide an estimate for the deviation from special cases of pure annihilations. For example, we observe that there is a special group of $\sim50$ models with masses $m_{\LSP}\approx 100\gev$ that are constrained much more tightly than even the models annihilating purely to $b\bar{b}$, \ie, the points falling below the red ``line" in the top left panel of \Figref{annihilations}. These are models (\cf, \Figref{shapes}) that have a significant enhancement of the $\gamma$-ray spectrum near the endpoint $x=E_{\gamma}/m_{\LSP}\approx1$. This enhancement is due to internal bremsstrahlung \cite{Bergstrom:1989jr}\cite{Bringmann:2007nk}, wherein an annihilation event produces a $\gamma$ ray directly from the hard process, \ie, $\LSP\LSP\to f\bar{f}\gamma$, in addition to radiation off of the final state particles created
 from the process $\LSP\LSP\to f\bar{f}$. The cross section for the
 $2\to3$ process is naively much smaller than that for the $2\to2$
 process as the $2\to3$ case is suppressed by a fine structure
 constant and by three-body phase space. However, the leading
 terms in the $2\to2$ process are helicity suppressed
 ($\sim(m_f/m_{\LSP})^2$) relative to the naive expectation, so that
 the $2\to3$ process, $\LSP\LSP\to f\bar{f}\gamma$, can be competitive
 with (or even dominate) the process $\LSP\LSP\to f\bar{f}$. This is
 especially true for annihilations to light final state particles
 $(m_f/m_{\LSP})\ll1$ through light superpartner mediators
 $(m_{\tilde{f}}/m_{\LSP})\approx1$. Here, we find that many models in
 this group have SUSY mass spectra with very light sleptons,
 $m_{\tilde{e}_R}\approx m_{\LSP}$ so that the would-be dominant
 annihilation channel $\LSP\LSP\to e^+e^-$ is heavily suppressed
 and the $2\to3$ channel $\LSP\LSP\to e^+e^-\gamma$ contributes
 greatly to the overall $\gamma$-ray spectrum.

\subsection{LSP Eigenstate Composition}

The mapping of colored points from the left panels to the right panels of \Figref{annihilations}, \ie, the combinations of $m_{\LSP}$, $\sigv$, $R$, and annihilation final-state distributions, is a complicated function of many of the couplings and masses that describe an arbitrary SUSY model. As such, it is difficult to robustly predict $\bigd$ for a given model by using only information about the interaction eigenstate composition of the lightest neutralino (our LSP DM candidate). While LSP eigenstate composition certainly has a large impact on the resulting $\gamma$-ray annihilation signal, it is difficult to disentangle the entire story from this piece of information alone.

\begin{figure}[t]
    \centering
    \includegraphics[width=1.0\textwidth]{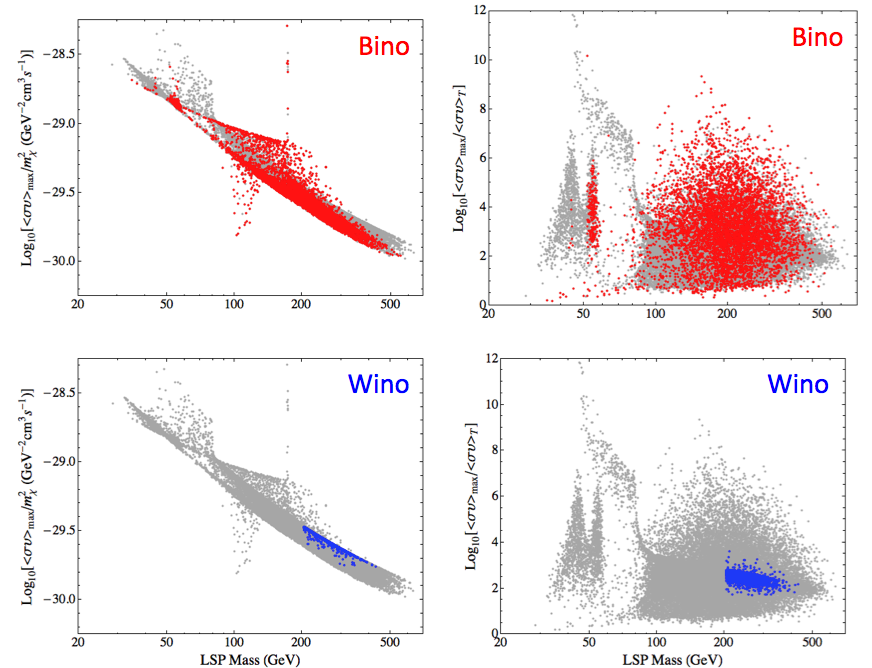}
    \caption{We display points in figures similar to those in \Figref{annihilations}. Here the full flat-prior model set is displayed as grey points and models categorized according to LSP eigenstate composition are overlaid in other colors. By convention our LSPs are described in terms of their neutralino mass matrix entries as: $\LSP=Z_{11}\tilde{B}+Z_{12}\tilde{W}^3+Z_{13}\tilde{H}_1^0+Z_{14}\tilde{H}_2^0$. Bino models are defined as having $|Z_{11}|^2>0.99$ and are displayed here in red. Wino models are defined as having $|Z_{12}|^2>0.99$ and are displayed here in blue.}
    \label{figs:ino-annihilations1}
  \end{figure}

\begin{figure}[t]
    \centering
    \includegraphics[width=1.0\textwidth]{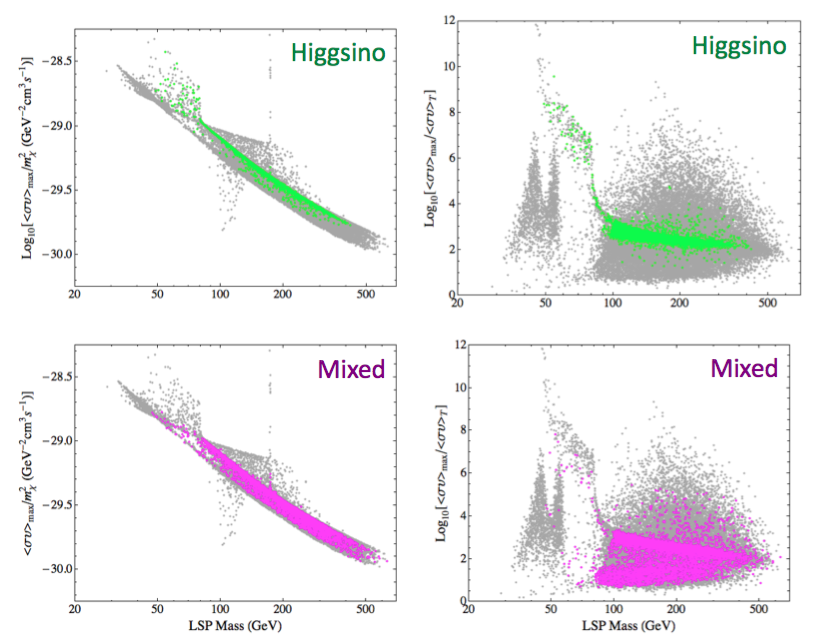}
    \caption{We display points in figures similar to those in \Figref{annihilations}. Here the full flat-prior model set is displayed as grey points and models categorized according to LSP eigenstate composition are overlaid in other colors. By convention our LSPs are described in terms of their neutralino mass matrix entries as: $\LSP=Z_{11}\tilde{B}+Z_{12}\tilde{W}^3+Z_{13}\tilde{H}_1^0+Z_{14}\tilde{H}_2^0$. Higgsino models are defined as having $(|Z_{13}|^2+|Z_{14}|^2)>0.99$ and are displayed here in green. Mixed models are defined as having $|Z_{11}|^2$, $|Z_{12}|^2$ and $(|Z_{13}|^2+|Z_{14}|^2)$ all $<0.7$ and are displayed here in magenta.}
    \label{figs:ino-annihilations2}
  \end{figure}

The impact of LSP eigenstate composition on the $\gamma$-ray signal is shown in Figures~\ref{figs:ino-annihilations1}-\ref{figs:ino-annihilations2}. We display points similarly as in \Figref{annihilations} except that they are now highlighted in color according to whether their LSPs are dominantly bino, wino, higgsino or mixed (with purities as denoted in the caption). These figures show that, while the resulting correlations are not as tight for eigenstate composition (Figures \ref{figs:ino-annihilations1}-\ref{figs:ino-annihilations2}) as those for annihilation final-state (\Figref{annihilations}), trends still exist. 

Bino-like LSPs annihilate essentially only through channels $\LSP\LSP\to f\bar{f}$ via t-channel exchanges of sfermion partners $\tilde{f}$. Due to helicity suppression, bino-like models typically annihilate to some combination of $\tau^+\tau^-$, $b\bar{b}$ and $t\bar{t}$ (compare the red points on \Figref{ino-annihilations1} with the red/green/orange points in the top panels of \Figref{annihilations}). The resulting mixture of rates into various final states depends on the pattern of sfermion masses, which are scanned over in our model generation procedure. The mapping in \Figref{ino-annihilations1} from left- to right-panel is thus quite sensitive to scanned parameters, resulting in a wide variety of predictions for $\gamma$-ray annihilation signals from bino-like LSPs.\footnote{For a large number of models in our set the calculation of $\sigv_T$ is also complicated by co-annihilations \cite{Griest:1990kh}, wherein the relic density is set by an effective annihilation rate $\sigv_{eff}$ in the early Universe, which is very different from the $\sigv$ governing annihilations in current DM halos.}

Wino-like LSPs annihilate dominantly to $\LSP\LSP\to W^+W^-$ by exchange of a chargino $\chip$ that is almost degenerate in mass with the LSP.\footnote{For models with very purely wino LSPs, the lightest chargino is highly degenerate in mass with the LSP. Null searches for new charged stable particles performed by CDF \cite{Abe:1992vr} and D0 \cite{Abazov:2008qu} have been applied in our model generation procedure \cite{Berger:2008cq}, excluding models with stable charginos lighter than $\lsim206\gev$. This results in a lower limit on the blue (wino-like) points (defined as having $|Z_{12}|^2>0.99$) in \Figref{ino-annihilations1}.} These models are seen to correlate well with the case of annihilations purely to $W^+W^-$ (compare the blue points on \Figref{ino-annihilations1} with the blue points in the top panels of \Figref{annihilations}). 

Higgsino-like LSPs couple efficiently to MSSM Higgs particles and to the $Z^0$, annihilating through a multitude of channels regardless of the sfermion masses. When $m_{\LSP}>m_{Z}$, these models annihilate efficiently through $\LSP\LSP\to Z^0Z^0$ and through $\LSP\LSP\to W^+W^-$ via exchange of the second neutralino, $\tilde \chi_2^0$, and the chargino $\chip$, which are nearly degenerate in mass with the LSP. Essentially all pMSSM models with higgsino-like LSPs annihilate dominantly through a mixture of the $W^+W^-$ and $Z^0Z^0$ final-state channels. The ratio of rates into the $W^+W^-$ and $Z^0Z^0$ final-state channels is primarily controlled by phase space (and asymptotes to $\sigv_{ZZ}/\sigv_{WW}\approx0.75$ for $m_{\LSP}\gg m_W\rm{,}m_Z$).\footnote{Higgsino models with $m_{\LSP}<m_W\mathrm{,}m_Z$ typically annihilate mostly into $\tau$-pair or $b\bar{b}$ final states although we observe a number of models where the $\gamma\gamma$ and $\gamma Z^0$ final states (proceeding through loops involving relatively light charginos) can become the dominant annihilation channels when the other relevant new particle masses are very heavy.} The fact that these models annihilate to a mixture of $W^+W^-$ and $Z^0Z^0$ final-state channels is reflected in the distribution of green points in \Figref{ino-annihilations2}. 

In our model set, models with mixed LSP compositions are seen to essentially always have a substantial higgsino fraction and are generally described as a bino-higgsino mixture with a sub-dominant wino component. Given this, the mixed LSP distributions on \Figref{ino-annihilations2} are unsurprising.

We observe that bino and mixed LSPs are over-represented in the set of models with $\bigd<10$, as compared to their occurrence in the full model set. This is to be expected, since many such models are seen to achieve a large LSP relic density $R\sim1$ while also maintaining a large annihilation cross section. In contrast, we find that \emph{none} of the models in the $\bigd<10$ have LSPs that are classified as nearly pure winos or higgsinos. This is also in accord with our expectations as such purely wino or higgsino LSPs annihilate too efficiently in the early Universe, leaving $R \ll 1$, at least for LSP masses below $\sim2-3 \TeV$.

\subsection{Comparison to Direct Detection}

It is expected that LHC searches, direct detection experiments, and indirect-detection experiments will provide highly complementary information about the nature of DM. The set of pMSSM models discussed in this work has already been studied in the context of LHC searches \cite{Conley:2010du}\cite{Conley:2011nn}, direct detection experiments \cite{Cotta:2009zu}\cite{Cotta:2011ht} and indirect detection searches with cosmic-ray electrons and positrons \cite{Cotta:2010ej} and neutrinos \cite{Cotta:2011ht}. We note that, essentially by construction, LHC searches are expected to rapidly exclude (or discover) most of the models in this set. Thus, a comparison of LAT and LHC results would be relatively unenlightening, since the most constraining LHC searches are typically only indirectly related to the detailed nature of the LSP. Therefore, we focus on a comparison between the prospects for future LAT dSph analyses and the limits on spin-independent and spin-dependent scattering cross sections. 

\begin{figure}[t]
  \centering
  \includegraphics[width=1.0\textwidth]{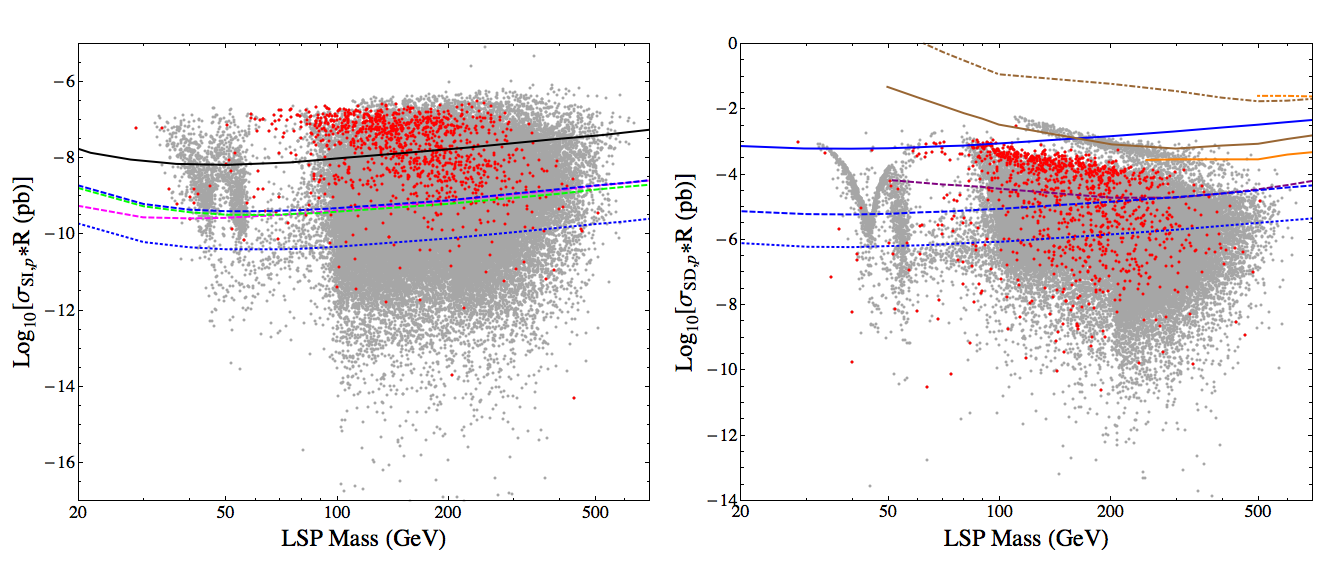}
  \caption{Comparison of LAT dwarf search and direct detection limits. We display all points in the flat-prior model set in grey and points having $\bigd<10$ in red. In the left panel the black curve depicts the current best SI scattering limit set by XENON100 \cite{Aprile:2010um}. Near-future projected SI limits from LUX \cite{McKinsey:2010zz}, SuperCDMS \cite{Bruch:2010eq}, COUPP 60kg and COUPP 500kg \cite{bigcoupp} are displayed as, magenta-dashed, green-dashed, blue-dashed and blue-dotted lines, respectively. In the right panel Current SD scattering limits from the AMANDA \cite{Braun:2009fr} and IceCube-22 \cite{Abbasi:2009uz} collaborations are displayed as brown and orange lines, respectively (with the assumption of soft or hard channel annihilations represented by dash-dotted or solid lines, respectively). Near-future projected experimental limits from the COUPP \cite{Behnke:2010xt}\cite{bigcoupp} 4kg, 60kg and 500kg searches in blue- solid, dashed and dotted lines, respectively. The projected IceCube/DeepCore limit estimated in \cite{Wiebusch:2009jf} is displayed as a magenta-dashed line (a more accurate IceCube/DeepCore analysis is presented in \cite{Cotta:2011ht}). 
  }
  \label{figs:dd}
\end{figure}

Since scattering signals are proportional to a single factor of the local dark matter density, direct detection experiments generally have an easier time setting limits on LSPs with low relic density   (scaling like $\sim R$ as opposed to $\sim R^2$ for indirect detection). In \Figref{dd}, we display the set of pMSSM models in the spin-independent (``SI,'' left panel) and spin-dependent (``SD,'' right panel) scattering cross section vs.\ LSP mass planes, highlighting the models within reach of future LAT dSph analyses (\ie, models with $\bigd < 10$). We note that spin-independent scattering bounds have become significantly more constraining since the era when the pMSSM models were generated. The current best bound has been set by the XENON100 Collaboration \cite{Aprile:2010um} and is depicted by the black curve on the left panel of \Figref{dd}. An uncertainty of about a factor of four applies to this curve, due to uncertainty in the determination of matrix elements for nuclear scattering.

We observe that there are many models that are expected to be discovered or excluded by both direct and indirect detection experiments. This is a fortunate scenario, potentially allowing for relationships between LSP mass, $\sigv$, $R$, annihilation final state channels and even details about heavier SUSY particles to be inferred. Additionally, we observe that there exist a number of models that will only be accessible to the LAT. These are models whose LSPs are dominantly bino and whose particle spectrum is somewhat hierarchical, including the light bino and one or more light sleptons. Such a scenario is essentially invisible in both direct detection experiments and at the LHC, due to a lack of accessible colored production channels. Generation of a new pMSSM model set that reflects progress in direct detection limits and the early running of the LHC is currently underway \cite{newmods}.

  \section{Conclusions}
  \label{sec:discuss}
We investigated the ability of the LAT instrument on board the \Fermi observatory to detect supersymmetric dark matter. LAT observations of ten Milky Way dwarf spheroidal satellite galaxies were combined to constrain the $\gamma$-ray signal expected from a large set of phenomenologically-viable SUSY models. The LAT analysis sets some of the tightest constraints on annihilation cross section yet available from indirect detection experiments; however, it falls slightly short of ruling out any of the SUSY models in our set. We note that many models are quite close to being excluded (or discovered) by the LAT. We expect that future LAT observations and improved instrument performance will put significant pressure on this region of parameter space.

We found that the majority of models in our set have a SUSY-dependence in accord with intuition developed from highly constrained scans of the MSSM. However, many models have unique annihilation channels and rates that are only observed in the broader context of the pMSSM. We have investigated the relative ability of the LAT to constrain annihilations into various final-state channels, noting that the LAT analysis is most sensitive to light LSPs ($m_{\LSP}<50\gev$) annihilating dominantly to $\tau$-pairs and heavier LSPs annihilating dominantly to $\bar{b}b$. Such behavior reflects a trade-off between the relative ease of constraining the spectral \emph{shape} of annihilations to $\tau$-pairs with the relatively low number of $\gamma$ rays that are produced as a result of these annihilations. The sensitivity crossover point is expected to move toward higher energies as the LAT continues to take data.

We have compared future expectations of the LAT dwarf search with those for direct detection experiments, finding examples of models that are accessible to combinations of the two experiment classes. Although the LAT search seems to be the more challenging method of discovering SUSY DM, we emphasize the unique character of such searches.  Indirect searches are invaluable due to their sensitivity to DM signals regardless of the SM states that the DM couples to most strongly (rather than requiring a strong coupling to quarks and gluons as in direct detection experiments and, to a large extent, colliders). Additionally, the indirect detection of DM would provide information about the cosmological DM abundance, rather than having to infer DM properties from neutral detector-stable particles produced at colliders. As the LAT mission continues, we expect it to extend the sensitivity of indirect searches for DM into this very interesting parameter space.

\acknowledgments
RCC gratefully acknowledges the support of the 2011 TASI Summer School, during which a part of this work was completed. ADW is supported in part by the Department of Energy Office of Science Graduate Fellowship Program (DOE SCGF), made possible in part by the American Recovery and Reinvestment Act of 2009, administered by ORISE-ORAU under contract no. DE-AC05-06OR23100.

The \textit{Fermi} LAT Collaboration acknowledges generous ongoing support from a number of agencies and institutes that have supported both the development and the operation of the LAT as well as scientific data analysis. These include the National Aeronautics and Space Administration and the Department of Energy in the United States, the Commissariat \`a l'Energie Atomique and the Centre National de la Recherche Scientifique / Institut National de Physique Nucl\'eaire et de Physique des Particules in France, the Agenzia Spaziale Italiana and the Istituto Nazionale di Fisica Nucleare in Italy, the Ministry of Education, Culture, Sports, Science and Technology (MEXT), High Energy Accelerator Research Organization (KEK) and Japan Aerospace Exploration Agency (JAXA) in Japan, and the K.~A.~Wallenberg Foundation, the Swedish Research Council and the Swedish National Space Board in Sweden.

Additional support for science analysis during the operations phase is gratefully acknowledged from the Istituto Nazionale di Astrofisica in Italy and the Centre National d'\'Etudes Spatiales in France.

\bibliographystyle{JHEP}
\bibliography{bib}

\end{document}